\documentclass[vecphys]{svmult}

\usepackage{graphicx}        
\usepackage[bottom]{footmisc}

\newcommand\micron{\mbox{$\mu$m}}%
\newcommand{\kms}{\mbox{km\,s$^{-1}$}}

\def\arcsec{\hbox{~$\!\!^{\prime\prime}$}}
\def\deg{\hbox{$^\circ$}}
\newcommand{\egg}{CRL2688}
\newcommand\apj{{ApJ}}%
\newcommand\apjl{{ApJ}}%
\newcommand\aap{{A\&A}}%
\newcommand\pasp{{PASP}}%
\newcommand\procspie{{Proc.~SPIE}}%
\begin{document}
\title*{Resolving the Multiple Outflows in the Egg Nebula 
with Keck II Laser Guide Star Adaptive Optics}
\titlerunning{Multiple Outflows in the Egg Nebula} 
\author{D. Le Mignant\inst{1, 2} \and R. Sahai\inst{3}
\and  A. Bouchez\inst{1}
\and  R. Campbell\inst{1}
\and J. Chin\inst{1}
\and  M. van Dam\inst{1}
\and E. Johansson\inst{1}
\and S. Hartman\inst{1}
\and R. Lafon\inst{1}
\and J. Lyke\inst{1}
\and P. Stomski\inst{1}
\and D. Summers\inst{1}
\and P. Wizinowich\inst{1}} 
\institute{W. M. Keck Observatory, Kamuela, HI 96743, USA 
\texttt{davidl@keck.hawaii.edu}
\and University of California Observatories, CA 95062, USA
\and Jet Propulsion Laboratory, Pasadena, CA 91109, USA
}
%
%
\maketitle
\begin{abstract}
The Egg Nebula has been regarded as the archetype of 
bipolar proto-planetary nebulae, yet we lack a coherent model that can
explain the morphology and kinematics of the nebular and dusty
components observed at high-spatial and spectral resolution. 
Here, we report on two sets of observations obtained with 
the Keck Adaptive Optics Laser Guide Star: 
H to M-band NIRC2 imaging, and narrow bandpath K-band 
OSIRIS 3-D imaging-spectroscopy (through the H2 2.121\micron ~emission
line). While the central star or engine remains un-detected at all 
bands, we clearly resolve the dusty components in the central region
and confirm that peak A is not a companion star.  
The spatially-resolved spectral analysis provide kinematic information
of the H$_2$ emission regions in the eastern and central parts of the
nebula and show projected velocities for the H$_2$ emission 
higher than 100\kms. We discuss these observations against a possible 
formation scenario for the nebular components.
\end{abstract}
\section{Introduction}
\label{sec:1}
The Egg Nebula is a prototypical bipolar pre-planetary nebula (PPN):
in the optical, the direct light from the central star is obscured by
dust, while two lobes aligned along a ``polar axis'' scatter the starlight
towards the
observer \cite{sahai98a}.
Yet, at longer wavelengths, the Egg Nebula reveals a more complex
geometry: 
multiple CO outflows in the equatorial plane and at higher
latitude \cite{cox00}; dusty knots (peak A) and a dark 
lane at a PA=140\deg ~\cite{goto02}; and 
H$_2$ (2.12\micron) emission line spatially coincident with the obscuring
dust in the equatorial plane \cite{kastner04}.
The models presented in \cite{sahai98b} or \cite{goto02} don't 
explain some of these
features such as the existence of the multiple outflows.\\
We present some of the near-infrared imaging and imaging-spectroscopy
results at high spatial ($\leq$0.1\arcsec) and spectral ($\approx$60\kms) 
resolution from the Egg Nebula obtained with the Keck II 
Laser Guide Star Adaptive Optics (LGS-AO) System. 
These LGS-AO images of the central regions show details 
on the dusty knots and the H$_2$ emission with 
un-precedent spatial 
resolution. Combined with ACS/HST F606W images and LGS-AO 3-D imaging 
spectroscopy through the H$_2$ line, we analyze 
the detailed structure of the outflows for the eastern part of the
nebula. We tentatively propose that multiple highly collimated
outflows, launched by intermittent and precessing mechanism 
could explain the Egg morphology. A paper describing the full
NIRC2 imaging data set, compared to visible HST ACS and 
including an analysis of the 3-D imaging spectroscopy over 
all outflows is in preparation \cite{lemignant08}.
\section{NIRC2 imaging and OSIRIS imaging-spectroscopy}
\label{sec:2}
%
%
In July 2004, we used the AO-dedicated near infrared camera NIRC2 behind 
the Keck II Laser Guide Star 
(LGS) Adaptive Optics (AO) system \cite{wizinowich06}, 
\cite{vandam06}, to record broad-band 
and narrow-band images. Earlier LGS-AO observations of the Egg were 
reported in \cite{bouchez04} and suffered from variable image quality.
Later in 2006, we were allocated Director's Keck II
LGS time for imaging 
spectroscopy of the Eastern and Center regions of the 
\egg~ using OSIRIS, the AO-dedicated Integral 
Field Unit \cite{larkin06}. Table \ref{tab:1} summarizes the observations
presented here.
\begin{small}
\begin{table}
\centering
\caption{Keck LGS-AO observations of the Egg Nebula reported in this work}
\label{tab:1}   
\begin{tabular}{lccccc}
\hline\noalign{\smallskip}
Region & Instrument  &  Spatial sampling & Filter & $\lambda_0$ & $\Delta_{\lambda}$\\
\noalign{\smallskip}\hline\noalign{\smallskip}
5\arcsec$\times$5\arcsec centered  &  NIRC2 & 0.\arcsec01/pix   & Lp         & 3.78  & 0.70 \\
5\arcsec$\times$5\arcsec centered  &  NIRC2 & 0.\arcsec01/pix   & Ms         & 4.67  & 0.24 \\
10\arcsec$\times$10\arcsec centered  &  NIRC2 & 0.\arcsec01/pix  & Kp         & 2.12  & 0.35     \\
40\arcsec$\times$40\arcsec centered  &  NIRC2 & 0.\arcsec04/pix
& H$_{2 {\nu=1-0}}$  & 2.121 & 0.034    \\
40\arcsec$\times$40\arcsec centered  &  NIRC2 & 0.\arcsec04/pix   & Kcont      & 2.270 & 0.030   \\
4\arcsec$\times$6\arcsec centered &  OSIRIS & 0.\arcsec1/lenslet  & Kn2   & 2.089 & 0.105  \\
4\arcsec$\times$6\arcsec on east &  OSIRIS & 0.\arcsec1/lenslet   & Kn2   & 2.088 & 0.105 \\
\noalign{\smallskip}\hline
\end{tabular}
\end{table}
\end{small}
\subsection{Results from NIRC2 Imaging}
\label{sec:2.1}
A spatial resolution of 70 milli-arcsec was estimated from the NIRC2
wide-camera images.
All NIRC2 wide camera images were registered against archived
ACS WFC1 F606W observations from October 2002 (data set j8gh55010, 
P.I. R. Sahai)
%
\begin{figure}
\centering
\includegraphics[angle=0,height=5.5cm]{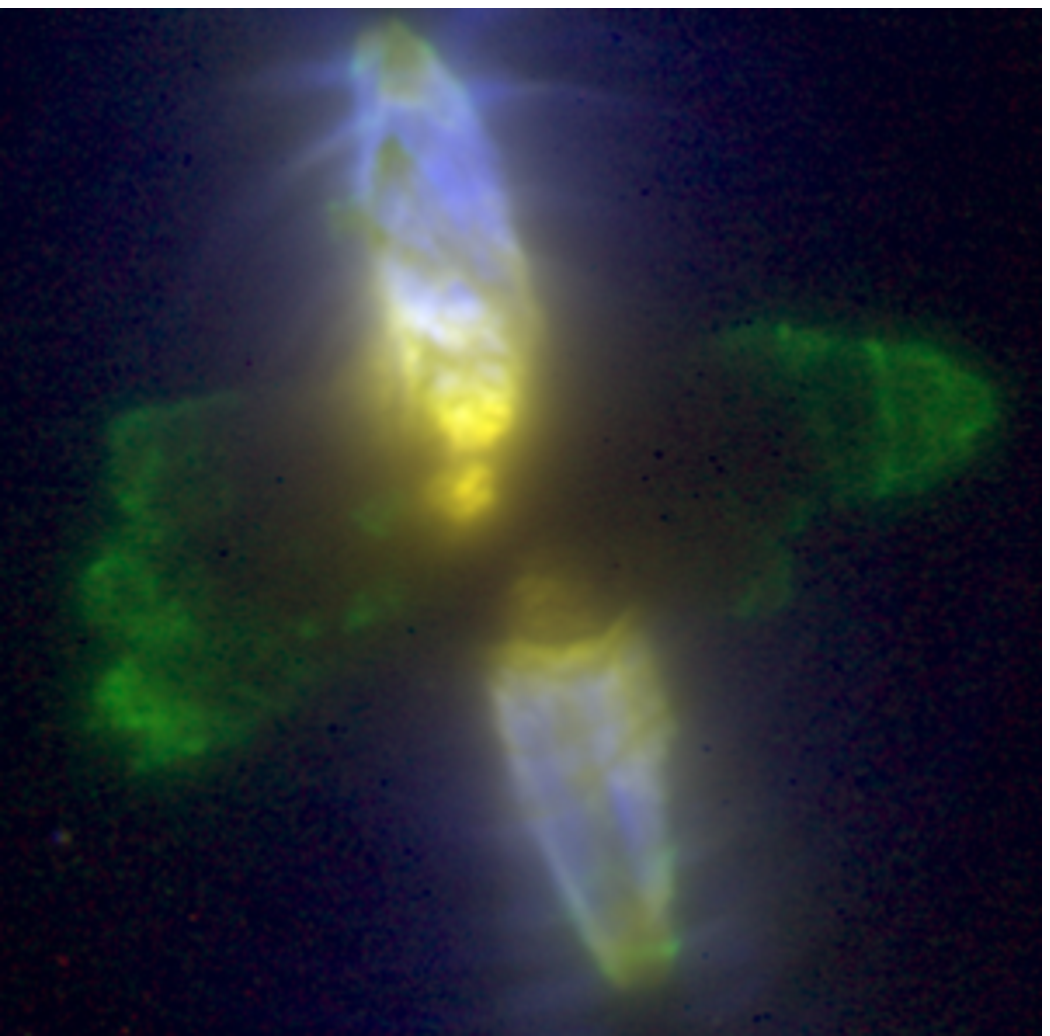}
\includegraphics[angle=0,height=5.5cm]{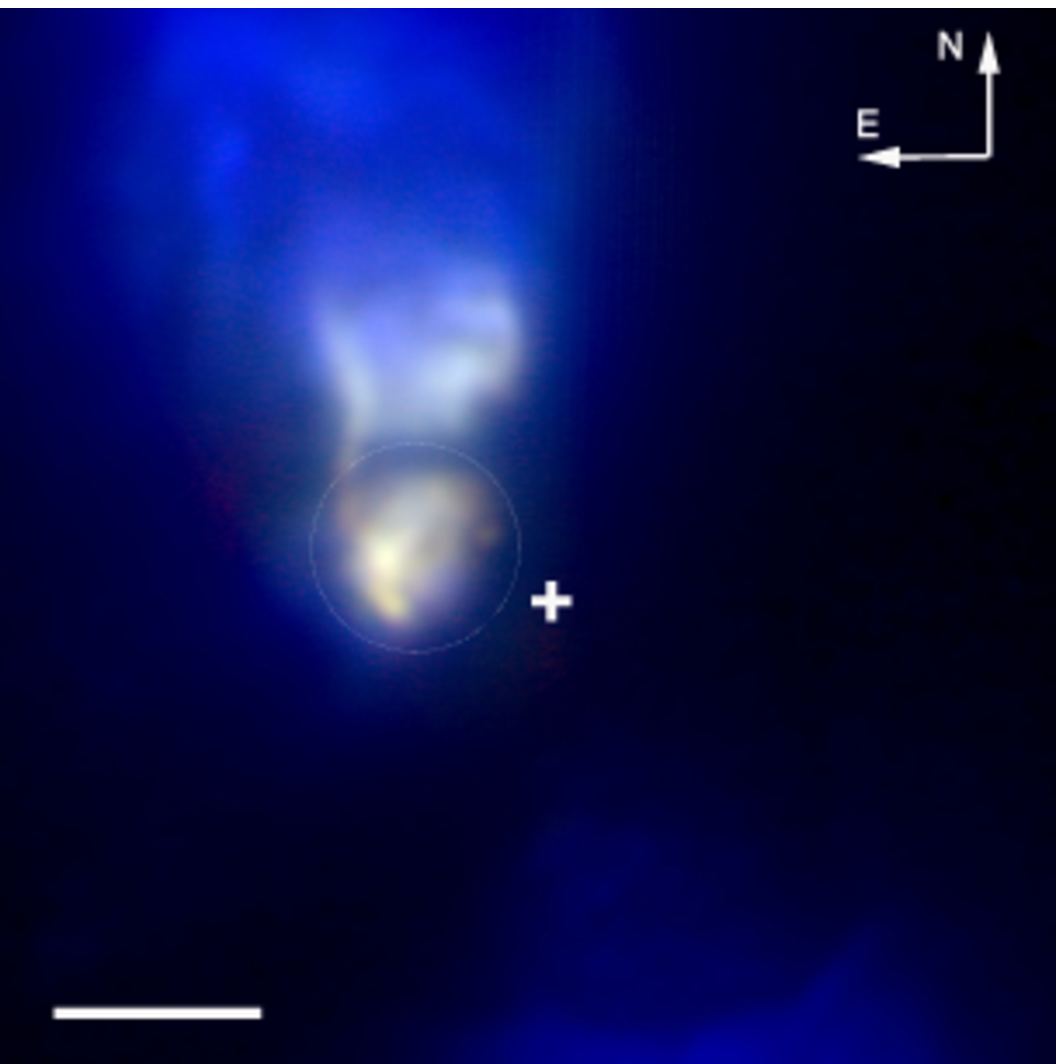}
\caption{{\bf Left:} 20\arcsec$\times$ 20\arcsec false-color image
  consisting of ACS F606W POL-0 (blue), NIRC2
  H$_2$ (green) and NIRC2 Kcont (red) images.  
 {\bf Right} 5\arcsec$\times$
 5\arcsec false-color NIRC2 image of the central region using  
Kp-band (blue), Lp-band (green) and Ms-band (red) images; 
The display 
intensity scale is I$^{0.25}$. 
The scale bar is 1\arcsec. The cross marks the central
star location, and a light circle surrounds peak A.}
\label{fig:1} 
\end{figure}
Fig.~\ref{fig:1} displays a color-combined NIRC2 and HST 
image and a color-combined NIRC2 image (see caption). 
There are a few noticeable features: 1) the visible (blue) light is
primarly scattered inside the two N and S lobes and along the two
searchlight beams \cite{sahai98a}; the scattered light in the 
K cont., Lp and Ms band is also seen along the N and S lobes and
closer to the central hidden star. 
Except for the N/S lobes, the only other location where scattered
light is seen in the NIR is around peak A
\cite{weintraub00}; there is no detection of K, Lp or Ms 
continuum emission in the equatorial region,
indicating a total absence of continuum scattered light in this region;
2) the H$_2$ emission regions appear as dense knots, filaments
 and arcs spatially coincident with the edges of the
obscuring/scattering dust knots seen along the N/S lobes, at the tips
of the outflows in the equatorial region, and at the tips of the
mid-latitude mulitple outflows, some of them, G$_1$, E$_1$,
D$_1$, reported in the lower resolution CO maps from \cite{cox00}; 
3) the 2004 Lp and Ms images
 show that peak A is now well-resolved into a 0.\arcsec5
dusty knot with a complex structure, possibly indicative of a nascent
outflow in the direction towards the observer; 
in the same image, there is a unresolved wiggling filamentary structure 
bridging from the N lobe onto peak A that could be
indicative of highly-collimated processes (e.g., a precessing jet) 
in the vicinity of the central engine.
\subsection{Results from OSIRIS Imaging Spectroscopy}
Two regions in the Egg nebula were observed using OSIRIS: one centered
on the central
star, the second one centered on the East H$_2$ emission region.
The lenslet choice of 0.\arcsec1 (in order to cover 
4\arcsec$\times$6\arcsec ~area at once) defines our spatial resolution.
The data was reduced using the OSIRIS Data Reduction
Pipeline.
%
%
\begin{figure}
\centering
\includegraphics[angle=0,height=4cm]{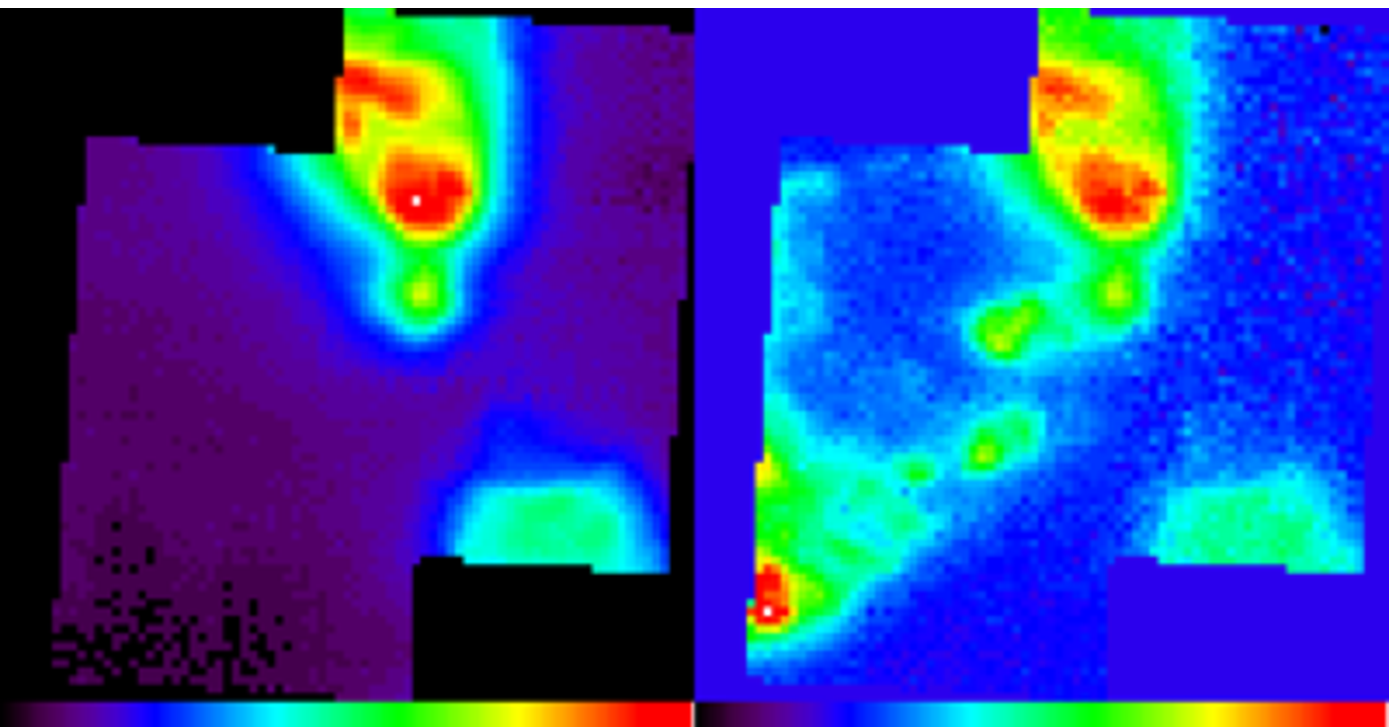}\\
\includegraphics[angle=0,height=4cm]{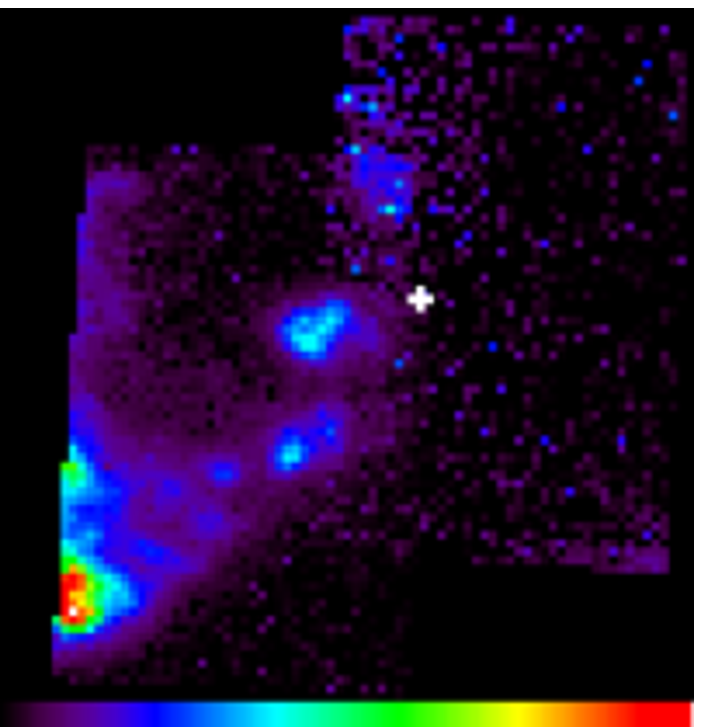}
\includegraphics[angle=0,height=4cm]{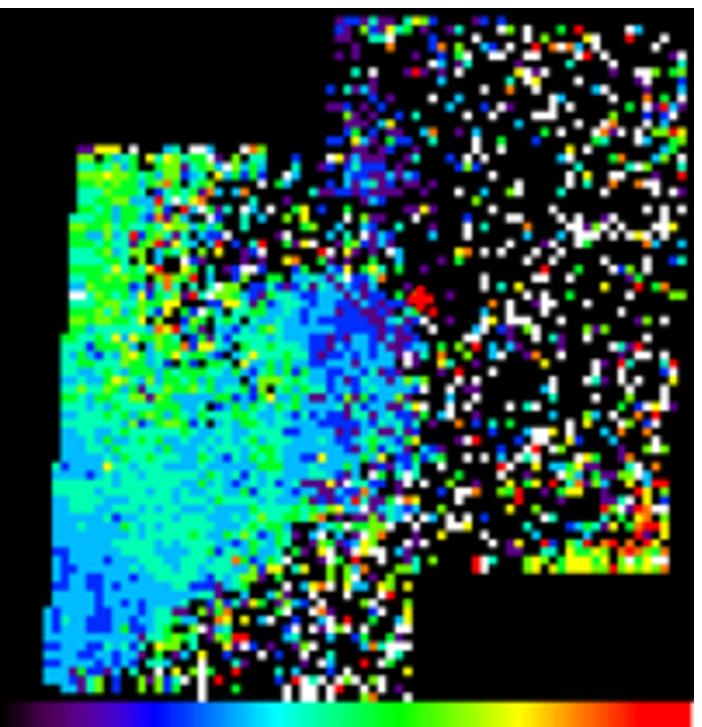}
\caption{{\bf Upper row}  
 The left mosaic image corresponds to the average over the 421
 spectral channels, therefore dominated by scattered light, 
while the right mosaic image is the average over
 the H$_2$ line (no continuum has been subtracted). The image
 intensity scale is I$^{0.25}$.
{\bf Bottom left} Map of the continuum-subtracted integrated 
flux for the shocked H$_2$ emission line (using linear intensity scale). 
{\bf Bottom right} Map of the H$_2$ line peak centroid shift in \kms  
~(using linear intensity scale). The velocity shift covers a range of
100~\kms.
 {\bf --} Each mosaic image is 8\arcsec$\times$8\arcsec 
with 0.\arcsec1 ~spatial resolution.}
\label{fig:2} 
\end{figure}
Fig.~\ref{fig:2} (upper row) 
shows a mosaic of the two 
regions observed, as the average over the cube or 
over the H$_2$ (2.12\micron) line.
A 421-channel spectrum for each 0.\arcsec1$\times$0.\arcsec1 ~spaxel 
over the 2.036--2.141\micron range was extracted and corrected for 
telluric absorption. The emission features include H$_2$ S(1) (1,0)
(2.1218\micron), H$_2$- S(3) (2,1) (2.073\micron) and HeI (2.1298\micron)
whereas the He I 4s 3S-3p 3P0 (2.1133\micron) is detected in absorption.\\ 
A spectral resolution of 60 \kms ~has been measured on the
unresolved telluric OH lines. For each spaxel, we fit a continuum and a 
gaussian profile through the H$_2$ emission line and
extracted relative integrated intensity, velocity dispersion
and line peak centroid.\\
The map of continuum-subtracted H$_2$ flux (see Fig.~\ref{fig:2} bottom
left) shows that the H$_2$ emission 
is not uniformly distributed over the observed region, but rather 
appears as 4 to 6 clumpy knots of angular size of $\approx$ 
1\arcsec ~or less, roughly aligned along an arc joining the East 
lobe to the N lobe.  None of these clumps is superimposed with peak A
location.\\
The shift in peak centroid is displayed on the
right-most image. We measure a velocity shift amplitude for this part
of the nebula of 100\kms, a factor 2 higher than previous
work \cite{kastner01}. The highest projected velocities are found 
near (but not at) peak A, at smaller spatial scales ($\leq$
0.\arcsec3) which could not be probed in earlier studies. The 
high-velocity
outflows north of the central star (cross) coincide 
with the clumpy H$_2$ knots closer to the N lobe and
could be associated with a precessing jet, suggested in ~\ref{sec:2.1}. 
\section{Discussion and conclusion}
From the preliminary analysis of our data sets, we conclude 
1) the absence of continuum light in most 
ouflows, 2) a distribution of H$_2$ emission coincident with dense, dusty 
clumps, 3) a possible nascent outflow at the peak A location,
and 4) the existence of fast highly-collimated ouflows, 
possibly jets.
We stress that any qualitative model must account for the
non-detection of continuum light in the equatorial outflows, 
in contrast to the polar ones.
The continuum light from the central star must be heavily 
obscured either in the equatorial ouflow cavity or before 
it reaches the cavity.
We speculate
a possible shaping mechanism that could account for these
observables: 
the central engine of the Egg is 
heavily obscured by a cocoon of
dust and intermittent jet-like outflows may be launched by 
MHD mechanisms within the vicinity of the central
engine \cite{blackman07}. 
An outflow would break through the dust cocoon, heating and 
dissipating the surrounding dust and expands for a few  hundreds of 
years \cite{ueta06}. 
The continuum starlight would propagate through the cavity and be
scattered. H$_2$ shocked 
emission would be
detected primarly at the tip of the outflow. 
Eventually, the 
continuum starlight becomes heavily obscured by dust again. 
In addition, new outflows would be launched rapidly and account for the 
multiple outflows with comparable dynamical age \cite{cox00}. 
The outflows where only H$_2$ shocked emission
is detected would correspond to past outflows.\\
We are currently completing LGS high-spatial 
imaging spectroscopy 
over the entire nebula to further constrain the kinematics 
and discuss the possible launch mechanism for the outflows.
%



\end{document}